\documentclass[twocolumn]{aa}
\usepackage{graphicx}
\usepackage{txfonts}
\begin{document}
\title{The rotational broadening and the mass of the donor star 
of GRS~1915+105}
\author{E.T. Harlaftis
          \inst{1}
          \and
	J. Greiner\inst{2}
	}
   \offprints{E. T. Harlaftis}
   \institute{Institute of Space Applications and Remote Sensing,
National Observatory of Athens, P. O. Box 20048, Athens 118 10, Greece\\
              \email{ehh@space.noa.gr}
	\and
	Max Planck Institute for Extraterrestrial Physics, Giessenbachstr., 
	PF 1312, 85741 - Garching, Germany\\
              \email{jcg@mpe.mpg.de}
	}
   \date{Received September 29, 2003; Accepted December 11, 2003 }

\abstract{
The  binary parameters   of  the microquasar  GRS~1915+105 have   been
determined by the detection of Doppler-shifted $^{12}$CO and $^{13}$CO
lines in its K-band spectrum (Greiner et al., 2001, Nature, 414, 522).  Here,
we present further  analysis  of the  same K-band  VLT spectra  and we
derive  a rotational broadening  of the donor   star of  $\upsilon \sin \
i=26\pm3$ km s$^{-1}$ from  the  $^{12}$CO/$^{13}$CO lines.   Assuming
that   the K-type star  is  tidally locked  to  the black  hole and is
filling its Roche-lobe surface, then the implied mass ratio is $q =
\frac{M_{\rm d}}{M_{\rm x}} =  0.058\pm0.033$.  This result,  combined
with ($P, K, i$)=(33.5 d,  140 km s$^{-1}$, 66$^{\circ}$) gives  a
more  refined   mass estimate   for   the       black    hole, $M_{\rm
x}=14.0\pm4.4~M_{\odot}$,    than previously   estimated,    using  an
inclination of  $i=66^{\circ}\pm2^{\circ}$  (Fender  et al.  1999)  as
derived  from the orientation  of the radio  jets  and a more accurate
distance. The  mass for  the   early K-type   giant  star  is  $M_{\rm
d}=0.81\pm0.53~M_{\odot}$,  consistent      with a  more       evolved
stripped-giant donor star in GRS~1915+105 than, for example, the donor 
star of the prototype black-hole X-ray transient, V404  Cyg
which has the longest binary period after GRS~1915+105.

}
   \maketitle

\section{Introduction}

The microquasar   GRS~1915+105  remains one  of  the  most  exotic and
variable objects in  the X-ray, IR and  radio spectrum bands since its
discovery (Castro-Tirado et al. 1994). A major accomplishment has been
the  implied  connection between  the jet   and  the accretion disk as
interpreted through modelling  of its  X-ray  variability (Belloni  et
al.  1997;  Klein-Wolt et al.  2002).  Thereafter,  it is evident that
crucial   insight into the accretion  physics  of jet formation can be
derived   through      intense  theoretical       and    observational
work.   Fundamental for the modelling  but   also for the evolutionary
history of GRS~1915+105 are the binary system parameters.

The binary model of the   microquasar GRS~1915+105 was elusive,  until
recently, despite many observational  efforts to search for the  donor
star (Castro-Tirado et al.  1996; Eikenberry  et al.  1998; Mirabel et
al.  1997; Mart\'{\i} et  al. 2000; Harlaftis et  al. 2001; Greiner et
al.    2001a).    Greiner  et   al.  (2001b)    were   able  to obtain
phase-resolved IR   spectra  using the   VLT-Antu  with the   ISAAC IR
spectrograph and derive the  orbital period of  the binary system from
the detected Doppler-shifted $^{12}$CO and $^{13}$CO absorption bands.
The deduction  of the masses of  the  binary, using  the mass function
equation,

\[ f(M_{\rm x}) = \frac{P~K_{\rm d}^{3}}{2\pi G} =
\frac{M_{\rm x}~\sin^{3}i}{(1+q)^{2}} , \]
requires the   orbital period $P$,  the semi-amplitude   of the radial
velocity curve of the donor   star around the black-hole $K_{\rm  d}$,
the binary inclination $i$ and the mass ratio $q=M_{\rm d}/M_{\rm x}$,
where $M_{\rm d}$ is the mass of the donor star and $M_{\rm x}$ the mass
of the black-hole.  Indeed, the  orbital period and the semi-amplitude
of the radial velocity  curve of the  donor star around the black-hole
have been determined by  Greiner  et al.  (2001b; $P=33.5\pm1.5$  days
and $Kw_{\rm  d}=   140\pm15$  km  s$^{-1}$).   The  remaining  unknown
parameters are the binary inclination and the mass ratio.  In fact, it
is generally the binary inclination which dominates the uncertainty in
the  mass  determination  of  the  black  hole via    $ \sim  f(M_{\rm
x})~\sin^{-3}i$.  The general method   in deriving the inclination  of
X-ray  binaries and cataclysmic  variables is to model the ellipsoidal
modulations  of the companion star at  infrared wavelengths.  Luckily,
the orbital inclination of  GRS~1915+105 is deduced with unprecedented
accuracy from the orientation of the jets ($i=70^{\circ}\pm2^{\circ}$;
Mirabel  \&  Rodriguez  1994;  $i=66^{\circ}\pm2^{\circ}$ by Fender et
al. 1999).   The final   parameter   that is  still  undetermined  for
GRS~1915+105 is its mass ratio.   The mass ratio  of the binary system
is related to  the rotational broadening of the  donor star  using the
formula

\[ \frac{\upsilon \sin i}{K_{\rm d}} =
0.46 \left[ (1+q)^{2} ~q \right] ^{1/3}  \]
where   $\upsilon  \sin  i$ is   the   rotational  broadening in   the
line-of-sight, assuming Roche  lobe mass-transfer, spherical  geometry
for the star and tidal  locking to the  primary star (Gies and  Bolton
1986;  Horne, Wade, Szkody 1986).  Here,  we further analysed the same
IR spectra from VLT   (Greiner et al. 2001b)  in  order to deduce  the
rotational  broadening  of the  donor   star from   the width  of  the
photospheric   absorption  lines  (mainly  $^{12}$CO, $^{13}$CO;   see
Greiner et al. 2001a).

\section[]{Observations, numerical technique and deduction of mass ratio}

The GRS~1915+105 spectra, together with a  KIII spectrum (HD~ 138185),
were obtained between     April   and August  2000 with     the  ISAAC
spectrograph on the Antu-VLT covering the range 2.2896--2.4131 $\mu$m, \ at
a pixel resolution  of 15.4 km  s$^{-1}$  (Greiner et  al. 2001b).  In
order  to measure the width of  the photospheric  absorption lines, we
apply a numerical  technique according to which  we subtract from  the
Doppler-corrected spectrum of  GRS~1915+105 a KIII  template spectrum.
In detail

\begin{itemize}
\item we produce the average spectrum after the individual spectra are Doppler-corrected
 by applying  the radial velocity solution (Greiner et al. 2001b)
\item we apply to the KIII template spectrum a 
rotational profile with a limb darkening coefficient of 0.5 (Gray 1992)
and a veiling factor $f$ which scales the depth of the absorption lines 
in order to simulate the GRS~1915+105 absorption lines
\item the difference spectrum represents mainly the accretion disk spectrum with all the photospheric
   absorption bands arising from the KIII donor star of GRS~1915+105 having been subtracted
\item we perform a $\chi^{2}$-test between the difference spectrum 
and its smoothed version (by applying a Gaussian profile of a large  
FWHM=450 km s$^{-1}$ to make it flat)
which is a measure of how good the absorption lines of the donor star have been subtracted
for each rotational profile and scaling factor $0<f<1$ applied 
\item we iterate by changing the rotational profile width and the depth
of the absorption lines - using a scaling factor $f$ representing the 
the donor star line flux - 
until we reach $\chi^{2}$ values moving along a parabolic surface.
 
\end{itemize}

\begin{figure}
   \centering
   \includegraphics[angle=-90,width=9cm]{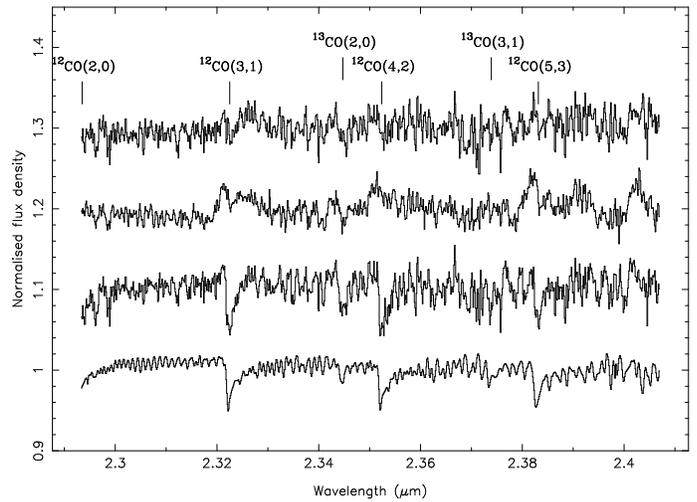}
\caption{The spectra show, from bottom to top, the KIII template 
(with lines multiplied by
0.14), the  average  Doppler-corrected  spectrum  of GRS~1915+105  (16
spectra), the difference spectrum (thus, representing  the  disk
spectrum) and the difference spectrum   between the template and   the
average Doppler-corrected  spectrum of  GRS1915+105  at binary   phase
$\sim$0.75 (the same   average spectrum as  in   Greiner et al.   2001b).}
\label{avdiff}
\end{figure}

The minimum $\chi^{2}$ on the parabola  ($\upsilon \sin i, ~\chi^{2}$)
gives  the corresponding width of  the rotational profile. We deduce a
$\upsilon \sin  i=26\pm3$ km s$^{-1}$  and $f=14\pm3$  \%, at the 90\%
significance level, where $f$ is the light contribution from the donor
star as estimated from its $^{12}$CO/$^{13}$CO absorption bands.  As a
validation  test,  we performed the same  iterations  but now  for the
individual spectra rather than  the average Doppler-corrected spectrum
and took the average $\upsilon \sin i$ - equal to $26\pm5$ km s$^{-1}$
- from the   individual estimates.  The  mass ratio  deduced  is, only
weakly sensitive,  on the spectral  type and veiling factor $f$.  For more
detailed discussion of the technique  see Harlaftis et al. (1999 and 
references therein). Fig.  \ref{avdiff}
shows, from bottom to top, the KIII template (with lines multiplied by
the veiling factor of 0.14), 
the  average  Doppler-corrected  spectrum  of GRS~1915+105  (16
spectra), the     difference spectrum (thus,   representing  the  disk
spectrum) and the difference spectrum   between the template and   the
average Doppler-corrected  spectrum of  GRS~1915+105  at binary   phase
$\sim$0.75 (same   average as  in   Greiner et al.   2001b).  The KIII
spectrum has a ripple in the continuum  arising from variations in the
star brightness between    integrations (i.e. at  different   detector
positions).   There   is a  residual    blue-shifted CO-band or  other
unidentified emission in the disk spectrum from phases other than 0.75
as well  as  two  unidentified absorption features  around  2.2962 and
2.2988 $\mu$m.

\section{Discussion}

We  discuss now the various assumptions  involved in the derivation of
the  mass ratio    from  the rotational  broadening.    The absorption
vibration-rotation bands $^{12}$CO and $^{13}$CO consist of $\sim4500$
lines at 3500 K (Kunde 1977).   However the total  damping width is of
the order  of  $1 \times 10^{-4}$  \AA  ~, the thermal  width does not
exceed 0.04 \AA~ at 4000 K, and the  macroturbulence width is $\sim 5$
km s$^{-1}$ in  giant  stars (Gray 1992).  Other   possible broadening
factors other    than rotation is  pressure-broadening   which  is not
significant in giant stars. Thus, the fact  that there is a measurable
velocity   dispersion  in the  line    profiles indicates that  it  is
dominated by  the  rotational broadening profile.  Shahbaz  (2003) has
developed a new, more accurate method involving synthetic spectra from
the Roche lobe (thus avoiding assumptions  about a spherical symmetric
star with  a limb darkening  coefficient).  However,  simulations show
that the standard method compared  to  his more accurate treatment  is
still compatible within 3$\sigma$.  The assumption of a spherical star
in the ($\upsilon \sin i, q$) relation has  been shown to introduce an
error in the  evaluation of mass  ratio $q$ of  $<5$ \%  (Marsh et al.
1994)  which  is much  less than  the uncertainty   of our measurement
($\sim  50$~\%), hence with  a negligible  effect.   The assumption of
synchronous rotation ($P_{\rm orbit}=P_{\rm rot}$) has been  found to be 
valid
for binaries as  wide as 80 days even   though giant stars may  show a
weaker  synchronism (10\%  difference  from the  orbital  period) than
late-type  stars  (Giuricin et  al.   1984).  The  criterion for tidal
locking is $R_{\rm d}/\alpha > 0.125$, where $R_{\rm d}$ is the radius
of the  donor star    and $\alpha$  the  binary   separation, assuming
dynamical tide  alone  (Zahn 1977;  Zahn 1992) which  indeed holds for
GRS~1915+105   ($R_{\rm   d}/\alpha  =    0.176$).   Involving   other
dissipative processes such  as  hydrodynamical meridional flow  on the
star lowers  the limit for  the fractional  star radius, thus yielding
much longer  periods (Tassoul 1987;  Tassoul \& Tassoul 1992; see also
for  tidal interaction models Witte  \& Savonije 2002, Claret \& Cunha
1997, Verbunt  \&  Phinney 1995).   The synchronisation timescale  $ t
\approx (R_{\rm  d}/\alpha)^{-6} ~ q^{-2}=4.6$ million years (Giuricin
et al.  1984) which is well within  the nuclear evolution timescale of
the donor star. Synchronisation is more  rapid than circularisation of
the orbit  (Hall  1986;  see also  Lecar  et  al.   1976), and a
$\chi_{\nu}^{2}$-test   on   the  radial   velocity  data (Greiner  et
al. 2001b) does not justify an eccentric orbit ($e<0.006$). Therefore,
we conclude that the orbit is circular.

The constraint  on the  eccentricity  of the radial velocity  fit also
indicates that there is no  line irradiation affecting the measurement
of the $K$ semi-amplitude of the radial velocity of GRS~1915+105.  For
example,  a 10\% correction   is applied  in   the value  of the   $K$
semi-amplitude (reduced) from the NaI  doublet absorption lines of the
donor star   in the  eclipsing dwarf  nova   IP Peg  from   a measured
eccentricity of the radial  velocity fit of  ($e=0.089$; Martin et al.
1989).    There has been no   evidence  for irradiation affecting  the
photospheric absorption  lines  in   any  quiescent black-hole   X-ray
transients  with extreme mass ratios (e.g.    Harlaftis et al.  1996).
However,  outbursts    can   irradiate the donor     star   and affect
significantly the radial velocity amplitude,  as for example with  the
FIII-IV  sub-giant donor star in GRO  J1655-40  (Shahbaz et al.  2000)
which has a binary period of 2.6  days and a  large mass ratio of 0.42
(Shahbaz    2003).  The spectra  of  GRS~1915+105  were obtained during
relatively  quiet  states (no outburst  or  flare) and in addition its
binary orbit is  long with a small cross section of the donor
star for any irradiation to affect it. In fact, using the relation (2)
from Phillips et  al.  (1999) we find  the effect to  be only 2.5\% 
compared to the 11\% uncertainty in $K_{\rm d}$ for GRS~1915+105.

The mass  ratio thus derived from  the rotational  broadening width is
$q=0.058\pm0.033$ (see   second  relation in  the  Introduction).  The
masses   can  now  be  determined  from  $(P,K,i,q)=$($33.5\pm1.5$  d,
$140\pm15$   km s$^{-1}$, $66^{\circ}\pm2 ^{\circ}$,  $0.058\pm0.033$)
giving       $M_{\rm   x}=14.0\pm4.4~M_{\odot}$       and      $M_{\rm
d}=0.81\pm0.53~M_{\odot}$ (see  first relation in  the  Introduction).
The large black-hole mass uncertainty is  dominated by the uncertainty
in the semi-amplitude   radial velocity   of   15 km s$^{-1}$.     The
contributions from  the period uncertainty  and  mass ratio  are about
equal but only a few \% whereas that of the inclination uncertainty is
practically zero.  The large mass ratio  uncertainty of 50\% is due to
the  large  uncertainty  in  both  the  rotational broadening and  the
semi-amplitude  of the     radial   velocity ($\sim  10$\%  for   each
parameter).  Having made the above points for the assumptions involved
we  now discuss the  ambiguity that may  be  reside in the inclination
value.   The  jet axis may  not be  perpendicular to the  binary plane
though  a  constraint   is  placed to   be   less  than  $25\degr$  in
GRS~1915+105 (Maccarone 2002).   Indeed,  in one of the  microquasars,
namely  GRO J1655-40, the  binary inclination is  $70\degr$ (Greene et
al.  2001)  whereas  the  jet  inclination is  found  at  $85^{\circ}$
(Hjellming
\& Rupen   1995).  The proper motions  of  the  plasmoids ejected from
GRS~1915+105 have been found to vary in apparent velocity and this was
attributed to a change in the jet velocity (Rodriguez \& Mirabel 1999;
Fender et al.  1999).  However, it must be  noted that the alternative
explanation is that the jet axis may be precessing.  Indeed, Rodriguez
\& Mirabel (1999) mention a change by $\sim10^{\circ}$ of the ejection
axis within one month as well as a  10--20 \% difference in the proper
motion  measured by Fender  et  al (1999).  However, Maccarone  (2002)
concludes, on  the  basis  of close  agreement between   the dynamical
estimate and the resonance interpretation  of the observed X-ray QPO
frequencies  for  the black  hole  mass,  that a large
misalignment between the jet axis and the disk axis is ruled out in GRS~1915+105.

The above discussion must be taken  under consideration when using the
adapted  inclination  of   $i=66^{\circ}\pm2^{\circ}$ (Fender et   al.
1999) for a distance of $11.2\pm0.8$ kpc.  Assuming a $\sim10^{\circ}$
uncertainty  in the inclination due to  possible  jet axis precession,
the  masses   become   $M_{\rm  x}=11.6\pm3.3~M_{\odot}$  and  $M_{\rm
d}=0.68\pm0.43~M_{\odot}$   for  an    inclination  of  $i=76^{\circ}$
($M_{\rm      x}=16.9\pm5.9~M_{\odot}$            and          $M_{\rm
d}=0.98\pm0.65~M_{\odot}$  for  $i=56^{\circ}$).   However,  extensive
radio monitoring  of GRS~1915+105 over the  last 10  years indicates no
measurable   constant    precession,  thus   giving   ground   to  the
interpretation of varying jet   ejection   rather  than a  jet    axis
precession.  For direct  comparison  on  the improvement of  the  mass
estimate using the mass  ratio determined here,  we deduce a  $M_{\rm
x}=12.9\pm4.04~M_{\odot}$ compared to $M_{\rm x}=14.0\pm4.0~M_{\odot}$
for $i=70^{\circ}\pm2^{\circ}$ as adapted by Greiner et al. (2001a).


Single KIII stars have negligible $\upsilon
\sin  i$. Indeed, a  K1III  and a K5III giant star  have  a mass of  2.3, 2.2
$M_{\odot}$, a radius of 11, 28 $R_{\odot}$ and a $<\upsilon \sin i>$ of
2.5, and less than 1.5 km s$^{-1}$, respectively (Gray 1992).  Greiner
et  al.  (2001a) identified the luminosity  class of the donor star in
GRS~1915+105  as a giant star   mainly based  on the equivalent  width
ratio of the $^{12}$CO/$^{13}$CO  lines.   Using the Eggleton  formula
(1983) to determine  the Roche lobe size and  hereafter the density of
the  lobe-filling   star,  we  derive  $R_{\rm  d}=~19~R_{\odot}$  and
$\rho=2\times10^{-4}$ g cm$^{-3}$, using an orbital period of $P=33.5$
days and   a mass ratio  of $q=0.058$.    This   compares well with  a
single-giant   KIII1-5 classification   ($R  =~11-28~ R_{\odot}$   and
$\rho=13-4\times10^{-4}$ g cm$^{-3}$).  Thus,  the luminosity  class -
giant - indicates that the star's envelope can be contained within the
Roche lobe,  and thus evolutionary expansion  of the donor can sustain
Roche lobe mass transfer rate.  Wind-fed  accretion alone cannot power
the luminosities we observe from this system.

The long orbital period suggests that the evolution of GRS~1915+105 is
further     advanced   than V404   Cyg,   a   similar   system with  a
stripped-subgiant K0IV donor at a period of  6.5 days around the black
hole ($M_{\rm d} =  0.7^{0.3}_{0.2} M_{\odot}$; Shahbaz et  al. 1994).
The more evolved giant  star in GRS~1915+105 is  also indicated by the
smaller mass of the donor star  compared to V404  Cyg due to increased
mass loss sustained by nuclear evolution (King et al.  1996).  Indeed,
the mass of the donor star, a ``stripped-giant'', is lower by a factor
of three compared to a single giant star.  King  (1993) has shown that
the mass of the donor star can have  a maximum mass of $1.3 M_{\odot}$
while   its  actual mass depends    on the mass    of the helium core.
Adapting     the relation  $R_{\rm   d}   =  12.55  R_{\odot}  (M_{\rm
c}/0.25)^{5.1}$ from King (1993), we derive $M_{\rm c}=0.27 M_{\odot}$
for $R_{\rm d}=19~R_{\odot}$.   The  helium core  has  a mass  $M_{\rm
c}=0.27 M_{\odot}$,  one third of the remaining  mass of the star at $M_{\rm
d}=0.81 M_{\odot}$. More than 1 $M_{\odot}$ mass of the donor star has
already  been transferred to  the  black hole.  The orbital period  of
GRS~1915+105 is increasing and will reach at  least 45 days (section 3
in King 1993), a point when the  donor star will reach its helium-core
mass  and the mass-transfer  rate will  cease.   Actually, the stellar
size for the helium-core mass of $M_{\rm c}=0.27 M_{\odot}$ is $R_{\rm
d}=~18.6~R_{\odot}$  (relation (1) in  King 1993), suggesting that the
stellar size is  close to its possible  minimum, thus not far from the
time the mass transfer rate will cease.

GRS~1915+105  is in perpetual   outburst since its  discovery in  1992
whereas V404 Cyg has been in quiescence after its X-ray outburst. This
is  consistent with a higher mass   transfer rate expected towards the
end of  the nuclear evolution  of the  donor  star.  Indeed,  a higher
mass-transfer rate is expected in  GRS~1915+105 by 4--8 times compared
to V404 Cyg, by using  the relation, $  -\dot{M}_{2} \simeq 5.4 \times
10^{-9}(M_{\rm c}/0.25)^{7.11}  M_{\rm d}~ M_{\odot} $ yr$^{-1}$ (King
1993), compared to  an Eddington accretion-rate of $\dot{M}_{\rm Edd}
\simeq 1.5  \times  10^{-8}$ erg  s$^{-1}$   (for a 14  $M_{\odot}$
black-hole  mass). An increased orbital  period, for  example from the
6.5 days of V404 Cyg to the 33.5 days of GRS~1915+105, due to evolution
would result in longer outbursts and a decreased duty cycle as well as
higher super-Eddington accretion  rates  (King  et al.   2002).    
The observed X-ray   luminosity  during outburst  is
dependent on inclination and anisotropic radiation patterns, therefore
it is  not    easy to show  that    the accretion rate   is higher  in
GRS~1915+105.  The observed X-ray luminosity is super-Eddington at the
outburst maximum ($L_{\rm Edd} \simeq 1.8 \times  10^{39}$ erg
s$^{-1}$ for a  14 $M_{\odot}$ black hole   mass) in both V404  Cyg at
$L_{\rm x}({\rm max}) \simeq  3.75  \times 10^{39}(D/5~ {\rm  kpc})^{2}$ 
erg
s$^{-1}$ (Tanaka 1989) and in GRS~1915+105 at $L_{\rm x}({\rm max}) ~
\sim  5 \times
10^{39}(D/12~ {\rm kpc})^{2}$ erg s$^{-1}$ (Rau et al. 2003).
Interestingly, King  (2002)  recently proposed
that  GRS~1915+105, a  low-mass X-ray  binary  showing super-Eddington
accretion,  may  well be the   best  representative in  our  galaxy of
ultraluminous X-ray sources in elliptical galaxies.

It is clear that,  as far as the physical  mechanisms involved in  the
radiation emitted, this rich system will  be explored more efficiently
once the binary parameters    are  known more accurately.    A  better
sampling of  the orbital period will decrease  the  uncertainty in the
stars'  masses which     mostly arises   from   the  radial   velocity
uncertainty, as well  as the independent  determination  of the binary
inclination from the IR ellipsoidal  modulations for comparison to the
inclination  derived   from  the  radio jets   will  constrain further
evolutionary considerations of this peculiar system.

\begin{acknowledgements}
We thank Bart Willems for discussion on tidal interaction in close
binaries and an anonynous referee for helpful comments. 
The use  of   the MOLLY  software  package developed by
T. Marsh is acknowledged.
\end{acknowledgements}

\label{lastpage}


\begin{thebibliography}{99}
\bibitem[\protect\citeauthoryear{belloni}{1997}]{b1} 
Belloni T., M\'{e}ndez M., King A. R., van der Klis M., van Paradijs J.,
        1997, ApJ 479, L145
\bibitem[\protect\citeauthoryear{castro}{1996}]{cas2} Castro-Tirado A. J., Geballe T. R., Lund N., 1996, ApJ, 461, L99
\bibitem[\protect\citeauthoryear{castro}{1994}]{cas1}Castro-Tirado A. J., 
Brandt S., Lund N., Lapshov I., 
Sunyaev R. A., Shlyapnikov A. A., Guziy S., Pavlenko E. P., 1994, ApJS, 92, 469
\bibitem[\protect\citeauthoryear{claret}{1997}]{claret} Claret A., Cunha N.C.S., 1997, A\&A, 318, 187 
\bibitem{}Eggleton P. P., 1983, ApJ, 268, 368
\bibitem[\protect\citeauthoryear{eikenberry}{1998}]{e1} Eikenberry S. S., Matthews K., Morgan E. H., Remillard R. A.,
Nelson R. W., 1998, ApJ, 494, L61
\bibitem[\protect\citeauthoryear{Fender}{1999}]{f1} 
Fender R. P. et al. 1999, MNRAS, 304, 865
\bibitem[\protect\citeauthoryear{gies}{2001}]{g}Gies D. R., Bolton C. T., 
1986, ApJ, 304, 371
\bibitem{}Gray D. F., 1992, The observation and analysis of stellar photospheres,  Cambridge Astrophysics Series, Vol. 20, Appendix B, p. 432
\bibitem{} Greene J., Bailyn C. D., Orosz J. A., 2001, ApJ, 554, 1290
\bibitem[\protect\citeauthoryear{Greiner et al.}{2001a}]{g1} 
Greiner J., Cuby J. G., McCaughrean M. J., Castro-Tirado A. J., Mennickent R. E., 2001a, A\&A, 373, L37
\bibitem[\protect\citeauthoryear{Greiner et al.}{2001b}]{g2} Greiner J., Cuby J. G., McCaughrean M. J., 2001b, Nature, 414, 522
\bibitem[\protect\citeauthoryear{giuricin}{1984}]{gi1} Giuricin G., Mardirossian, Mezzetti M., 1984, A\&A, 141, 227
\bibitem[]{} Hall D. S., 1986, 309, L83 
\bibitem[\protect\citeauthoryear{harlaftis}{2001}]{har1} Harlaftis E. T., Dhillon V. S., Castro-Tirado A., 2001, A\&A, 369, 210
\bibitem[\protect\citeauthoryear{harlaftis}{1999}]{har2}
Harlaftis E. T., Collier S., Horne K., Filippenko A. V.,
1999, A\&A, 341, 491
\bibitem[\protect\citeauthoryear{harlaftis}{1996}]{har3}
Harlaftis E. T., Horne K., Filippenko A. V., 1996, PASP, 108, 762
\bibitem[\protect\citeauthoryear{hjellming}{1995}]{hje} Hjellming R. M., Rupen M. P., 1995, Nat, 375, 464
\bibitem[\protect\citeauthoryear{horne}{1986}]{h1} Horne K., Wade R. A., Szkody P., 1986, MNRAS, 219, 791
\bibitem[\protect\citeauthoryear{king}{2002}]{k1} King A.R., 2002, MNRAS, 335, L13
\bibitem[\protect\citeauthoryear{king}{1996}]{k3} King A.R., Kolb U., Burderi L., 1996, ApJ, 464, L127
\bibitem[\protect\citeauthoryear{king}{1993}]{k4} King A.R., 1993, MNRAS, 260, L5
\bibitem[\protect\citeauthoryear{klein}{2002}]{klein} Klein-Wolt, M., 
Fender, R. P., Pooley, G. G., Belloni, T., Migliari, S., Morgan, E. H.,
 van der Klis M., 2002, MNRAS, 331, 745
\bibitem[\protect\citeauthoryear{kunde}{1977}]{kun1} Kunde V. G., 1977, ApJ, 153, 435 
\bibitem[]{} Lecar M., Wheeler J. C., McKee C. F., 1976, ApJ, 205, 556
\bibitem[]{} Maccarone T. J., 2002, MNRAS, 336, 1371 
\bibitem[\protect\citeauthoryear{marsh}{1994}]{mar1} Marsh T. R., Robinson E. L., Wood J. H., 1994, MNRAS, 266, 137
\bibitem[\protect\citeauthoryear{marti}{2000}]{marti1} Mart\'{\i} J., Mirabel I. F., Chaty S., Rodr\'{\i}guez L. F., 2000, A\&A, 356, 943
\bibitem[\protect\citeauthoryear{Martin}{1989}]{martin1}
Martin J. S., Friend M. T., Smith R C., Jones D. H. P., 1989, MNRAS, 240, 519
\bibitem[\protect\citeauthoryear{mirabel}{1997}]{m2} Mirabel I. F., 
Bandyopadhyay R., Charles P. A., Shahbaz T., Rodriguez L. F., 1997,
ApJ, 477, L45
\bibitem[\protect\citeauthoryear{Mirabel}{1994}]{m1} Mirabel I. F., Rodriguez L. F., 1994, Nature, 371, 46
\bibitem{} Phillips S. N., Shahbaz T., Podsiadlowski P., 1999, MNRAS, 304, 839
\bibitem[\protect\citeauthoryear{rau}{2003}]{rau} Rau A., Greiner 
J., McCollough M. L., 2003, ApJ, 590, L37
\bibitem[\protect\citeauthoryear{rodriguez}{1999}]{r1} Rodriguez L. F., Mirabel F., 1999, ApJ, 511, 398
\bibitem[\protect\citeauthoryear{verbunt}{1995}]{v1} Verbunt F., Phinney E. S., 1995, A\&A, 296, 709
\bibitem[\protect\citeauthoryear{shahbaz}{2003}]{s1} Shahbaz T., 2003, MNRAS, 339, 1031
\bibitem[\protect\citeauthoryear{shahbaz}{2000}]{s2} Shahbaz T., 
Groot P., Phillips S. N., Casares J., Charles P. A., van Paradijs J., 
2000, MNRAS, 314, 747
\bibitem[\protect\citeauthoryear{shahbaz}{2000}]{s2} Shahbaz T., 
Ringwald F. A., Bunn J. C., Naylor T., Charles P. A., Casares J.,  
1994, MNRAS, 271, L10
\bibitem[\protect\citeauthoryear{tanaka}{1989}]{tan} Tanaka Y., 1989, 
The 23rd ESLAB Symposium on Two Topics in X-Ray 
Astronomy, Volume 1: X Ray Binaries, 3, ESA N90-25711-19-89
\bibitem[\protect\citeauthoryear{tassoul2}{1992}]{t2} Tassoul J.-L., Tassoul M., 1992, ApJ, 395, 259
\bibitem[\protect\citeauthoryear{tassoul}{1987}]{t1} Tassoul J.-L., 1987, ApJ, 322, 856
\bibitem[\protect\citeauthoryear{zahn2}{1992}]{z2} Zahn J.-P., 1992, in {\it Binaries as Tracers of Stellar Formation}, eds. A. Duquennoy, M. Mayor, Cambridge University Press, Cambridge, 252
\bibitem[\protect\citeauthoryear{zahn}{1977}]{z1} Zahn J.-P., 1977, A\&A, 57, 383
\bibitem[\protect\citeauthoryear{}{}]{} Witte M. G., Savonnije G. J., 2002, 
A\&A, 386, 222

\end{thebibliography}
\end{document}